\documentclass{aastex631}

\begin{document}

\title{A Multiwavelength Study of a Long-Duration VHE Flare from BL\,Lacertae with VERITAS}

\author[0000-0002-2028-9230]{A.~Acharyya}\affiliation{CP3-Origins, University of Southern Denmark, Campusvej 55, 5230 Odense M, Denmark}
\author{A.~Archer} \affiliation{Department of Physics and Astronomy, DePauw University, Greencastle, IN 46135-0037, USA}
\author[0000-0002-3886-3739]{P.~Bangale} \affiliation{Department of Physics, Temple University, Philadelphia, PA 19122, USA}
\author[0000-0002-9675-7328]{J.~T.~Bartkoske}\affiliation{Department of Physics and Astronomy, University of Utah, Salt Lake City, UT 84112, USA}
\author[0000-0003-2098-170X]{W.~Benbow}\affiliation{Center for Astrophysics $|$ Harvard \& Smithsonian, Cambridge, MA 02138, USA}
\author[0000-0001-6391-9661]{J.~H.~Buckley}\affiliation{Department of Physics, Washington University, St. Louis, MO 63130, USA}
\author[0009-0001-5719-936X]{Y.~Chen}\affiliation{Department of Physics and Astronomy, University of California, Los Angeles, CA 90095, USA}
\author{J.~L.~Christiansen}\affiliation{Physics Department, California Polytechnic State University, San Luis Obispo, CA 94307, USA}
\author{A.~J.~Chromey}\affiliation{Center for Astrophysics $|$ Harvard \& Smithsonian, Cambridge, MA 02138, USA}
\author[0000-0003-1716-4119]{A.~Duerr}\affiliation{Department of Physics and Astronomy, University of Utah, Salt Lake City, UT 84112, USA}
\author[0000-0002-1853-863X]{M.~Errando}\affiliation{Department of Physics, Washington University, St. Louis, MO 63130, USA}
\author{M.~Escobar~Godoy}\affiliation{Santa Cruz Institute for Particle Physics and Department of Physics, University of California, Santa Cruz, CA 95064, USA}
\author[0000-0002-4131-655X]{J.~Escudero~Pedrosa}\affiliation{Center for Astrophysics $|$ Harvard \& Smithsonian, Cambridge, MA 02138, USA}
\author[0000-0002-5068-7344]{A.~Falcone}\affiliation{Department of Astronomy and Astrophysics, 525 Davey Lab, Pennsylvania State University, University Park, PA 16802, USA}
\author{S.~Feldman}\affiliation{Department of Physics and Astronomy, University of California, Los Angeles, CA 90095, USA}
\author[0000-0001-6674-4238]{Q.~Feng}\affiliation{Department of Physics and Astronomy, University of Utah, Salt Lake City, UT 84112, USA}
\author[0000-0002-2636-4756]{S.~Filbert}\affiliation{Department of Physics and Astronomy, University of Utah, Salt Lake City, UT 84112, USA}
\author[0000-0002-1067-8558]{L.~Fortson}\affiliation{School of Physics and Astronomy, University of Minnesota, Minneapolis, MN 55455, USA}
\author[0000-0003-1614-1273]{A.~Furniss}\affiliation{Santa Cruz Institute for Particle Physics and Department of Physics, University of California, Santa Cruz, CA 95064, USA}
\author[0000-0002-0109-4737]{W.~Hanlon}\affiliation{Center for Astrophysics $|$ Harvard \& Smithsonian, Cambridge, MA 02138, USA}
\author[0000-0003-3878-1677]{O.~Hervet}\affiliation{Santa Cruz Institute for Particle Physics and Department of Physics, University of California, Santa Cruz, CA 95064, USA}
\author[0000-0001-6951-2299]{C.~E.~Hinrichs}\affiliation{Center for Astrophysics $|$ Harvard \& Smithsonian, Cambridge, MA 02138, USA and Department of Physics and Astronomy, Dartmouth College, 6127 Wilder Laboratory, Hanover, NH 03755 USA}
\author[0000-0002-6833-0474]{J.~Holder}\affiliation{Department of Physics and Astronomy and the Bartol Research Institute, University of Delaware, Newark, DE 19716, USA}
\author{Z.~Hughes}\affiliation{Department of Physics, Washington University, St. Louis, MO 63130, USA}
\author[0000-0002-1432-7771]{T.~B.~Humensky}\affiliation{Department of Physics, University of Maryland, College Park, MD, USA and NASA GSFC, Greenbelt, MD 20771, USA}
\author[0000-0002-1089-1754]{W.~Jin}\affiliation{Department of Physics and Astronomy, University of California, Los Angeles, CA 90095, USA}
\author[0009-0008-2688-0815]{M.~N.~Johnson}\affiliation{Santa Cruz Institute for Particle Physics and Department of Physics, University of California, Santa Cruz, CA 95064, USA}
\author[0000-0002-3638-0637]{P.~Kaaret}\affiliation{Department of Physics and Astronomy, University of Iowa, Van Allen Hall, Iowa City, IA 52242, USA}
\author{M.~Kertzman}\affiliation{Department of Physics and Astronomy, DePauw University, Greencastle, IN 46135-0037, USA}
\author{M.~Kherlakian}\affiliation{Fakult\"at f\"ur Physik \& Astronomie, Ruhr-Universit\"at Bochum, D-44780 Bochum, Germany}
\author[0000-0002-4260-9186]{T.~K.~Kleiner}\affiliation{DESY, Platanenallee 6, 15738 Zeuthen, Germany}
\author[0000-0002-4289-7106]{N.~Korzoun}\affiliation{Department of Physics and Astronomy and the Bartol Research Institute, University of Delaware, Newark, DE 19716, USA}
\author{S.~Kundu}\affiliation{Department of Physics and Astronomy, University of Alabama, Tuscaloosa, AL 35487, USA}
\author[0000-0003-4641-4201]{M.~J.~Lang}\affiliation{School of Natural Sciences, University of Galway, University Road, Galway, H91 TK33, Ireland}
\author[0000-0003-3802-1619]{M.~Lundy}\affiliation{Physics Department, McGill University, Montreal, QC H3A 2T8, Canada}
\author{E.~Meyer}\affiliation{Department of Physics, University of Maryland, Baltimore County, Baltimore MD 21250, USA}
\author{J.~Millis}\affiliation{Department of Physics and Astronomy, Ball State University, Muncie, IN 47306, USA and Department of Physics, Anderson University, 1100 East 5th Street, Anderson, IN 46012}
\author[0000-0001-5937-446X]{C.~L.~Mooney}\affiliation{Department of Physics and Astronomy and the Bartol Research Institute, University of Delaware, Newark, DE 19716, USA}
\author[0000-0002-1499-2667]{P.~Moriarty}\affiliation{School of Natural Sciences, University of Galway, University Road, Galway, H91 TK33, Ireland}
\author[0000-0002-3223-0754]{R.~Mukherjee}\affiliation{Department of Physics and Astronomy, Barnard College, Columbia University, NY 10027, USA}
\author[0000-0002-6121-3443]{W.~Ning}\affiliation{Department of Physics and Astronomy, University of California, Los Angeles, CA 90095, USA}
\author[0000-0002-4837-5253]{R.~A.~Ong}\affiliation{Department of Physics and Astronomy, University of California, Los Angeles, CA 90095, USA}
\author[0000-0003-3820-0887]{A.~Pandey}\affiliation{Department of Physics and Astronomy, University of Utah, Salt Lake City, UT 84112, USA}
\author[0000-0001-7861-1707]{M.~Pohl}\affiliation{Institute of Physics and Astronomy, University of Potsdam, 14476 Potsdam-Golm, Germany and DESY, Platanenallee 6, 15738 Zeuthen, Germany}
\author[0000-0002-0529-1973]{E.~Pueschel}\affiliation{Fakult\"at f\"ur Physik \& Astronomie, Ruhr-Universit\"at Bochum, D-44780 Bochum, Germany}
\author[0000-0002-4855-2694]{J.~Quinn}\affiliation{School of Physics, University College Dublin, Belfield, Dublin 4, Ireland}
\author[0000-0002-5104-5263]{P.~L.~Rabinowitz}\affiliation{Department of Physics, Washington University, St. Louis, MO 63130, USA}
\author[0000-0002-5351-3323]{K.~Ragan}\affiliation{Physics Department, McGill University, Montreal, QC H3A 2T8, Canada}
\author{P.~T.~Reynolds}\affiliation{Department of Physical Sciences, Munster Technological University, Bishopstown, Cork, T12 P928, Ireland}
\author[0000-0002-7523-7366]{D.~Ribeiro}\affiliation{School of Physics and Astronomy, University of Minnesota, Minneapolis, MN 55455, USA}
\author{L.~Rizk}\affiliation{Physics Department, McGill University, Montreal, QC H3A 2T8, Canada}
\author{E.~Roache}\affiliation{Center for Astrophysics $|$ Harvard \& Smithsonian, Cambridge, MA 02138, USA}
\author[0000-0003-1387-8915]{I.~Sadeh}\affiliation{DESY, Platanenallee 6, 15738 Zeuthen, Germany}
\author{A.~C.~Sadun}\affiliation{Department of Physics, University of Colorado Denver, Campus Box 157, P.O. Box 173364, Denver CO 80217, USA}
\author[0000-0002-3171-5039]{L.~Saha}\affiliation{Center for Astrophysics $|$ Harvard \& Smithsonian, Cambridge, MA 02138, USA}
\author{G.~H.~Sembroski}\affiliation{Department of Physics and Astronomy, Purdue University, West Lafayette, IN 47907, USA}
\author[0000-0002-9856-989X]{R.~Shang}\affiliation{Department of Physics and Astronomy, Barnard College, Columbia University, NY 10027, USA}
\author[0000-0003-3407-9936]{M.~Splettstoesser}\affiliation{Santa Cruz Institute for Particle Physics and Department of Physics, University of California, Santa Cruz, CA 95064, USA}
\author[0000-0002-9852-2469]{D.~Tak}\affiliation{SNU Astronomy Research Center, Seoul National University, Seoul 08826, Republic of Korea.}
\author{A.~K.~Talluri}\affiliation{School of Physics and Astronomy, University of Minnesota, Minneapolis, MN 55455, USA}
\author{J.~V.~Tucci}\affiliation{Department of Physics, Indiana University Indianapolis, Indianapolis, Indiana 46202, USA}
\author[0000-0002-8090-6528]{J.~Valverde}\affiliation{Department of Physics, University of Maryland, Baltimore County, Baltimore MD 21250, USA and NASA GSFC, Greenbelt, MD 20771, USA}
\author[0000-0003-2740-9714]{D.~A.~Williams}\affiliation{Santa Cruz Institute for Particle Physics and Department of Physics, University of California, Santa Cruz, CA 95064, USA}
\author[0000-0002-2730-2733]{S.~L.~Wong}\affiliation{Physics Department, McGill University, Montreal, QC H3A 2T8, Canada}

\collaboration{64}{(The VERITAS Collaboration)}

\author[0000-0003-2483-2103]{M. F. Aller}\affiliation{Department of Astronomy, University of Michigan, 323 West Hall, 1085 S. University Avenue, Ann Arbor, MI 48109, USA}
\author{M. Asadi-Zeydabadi}\affiliation{Department of Physics, University of Colorado Denver, Campus Box 157, P.O. Box 173364, Denver CO 80217, USA}
\author[0000-0003-1468-9526]{R. Hickox}\affiliation{Department of Physics and Astronomy, Dartmouth College, 6127 Wilder Laboratory, Hanover, NH 03755, USA}
\author[0000-0001-6158-1708]{Svetlana G. Jorstad}\affiliation{Institute for Astrophysical Research, Boston University, 725 Commonwealth Avenue, Boston, MA 02215, USA}
\author[0000-0001-6314-9177]{Sebastian Kiehlmann}\affiliation{Institute of Astrophysics, Foundation for Research and Technology-Hellas, GR-71110 Heraklion, Greece}\affiliation{Department of Physics, Univ. of Crete, GR-70013 Heraklion, Greece}
\author{P.~Lusen }\affiliation{Santa Cruz Institute for Particle Physics and Department of Physics, University of California, Santa Cruz, CA 95064, USA}
\author[0000-0001-7396-3332]{Alan P. Marscher}\affiliation{Institute for Astrophysical Research, Boston University, 725 Commonwealth Avenue, Boston, MA 02215, USA}
\author[0000-0002-5491-5244]{Walter Max-Moerbeck}\affiliation{Departamento de Astronom´ıa, Universidad de Chile, Camino El Observatorio 1515, Las Condes, Santiago, Chile}
\author[0000-0001-5957-1412]{P. V. de la Parra}\affiliation{CePIA, Astronomy Department, Universidad de Concepción, Casilla 160-C, Concepción, Chile}
\author[0000-0001-9152-961X]{Anthony C. S. Readhead}\affiliation{Owens Valley Radio Observatory, California Institute of Technology, Pasadena, CA 91125, USA}
\author{K. Riley}\affiliation{Department of Physics, University of Colorado Denver, Campus Box 157, P.O. Box 173364, Denver CO 80217, USA}
\nocollaboration{11}

\correspondingauthor{Claire E. Hinrichs}
\email{claire.hinrichs@cfa.harvard.edu}

\correspondingauthor{Atreya Acharyya}
\email{atreya@cp3.sdu.dk}

\begin{abstract}

We report the first observations of a long-duration very-high-energy (VHE; E$>$100 GeV) flare from BL\,Lacertae (VER J2202+422), taken with the Very Energetic Radiation Imaging Telescope Array System (VERITAS). On October 15, 2022, the \textit{Fermi}-Large Area Telescope (LAT)  detected elevated GeV activity originating from this blazar. This triggered a multiwavelength campaign, which includes observations from VERITAS, \textit{Swift}, \textit{NuSTAR}, and select optical and radio observatories. VERITAS observed the source for a total of $\sim$\,9.8 hours between September 1, 2022 and December 1, 2022. An analysis of these data yields a $\sim$\,28$\sigma$ detection of the source. While previously observed VHE flares from BL Lacertae have lasted on  time-scales of minutes to days, VERITAS continued to detect flaring activity from the source for over a month ($\sim$\,40 days) after the original flaring activity was detected with \textit{Fermi}-LAT. Broadband spectral modeling shows that a synchrotron self-Compton (SSC) model with an external inverse-Compton (EC) component is preferred over a one-zone SSC model.

\end{abstract}


\section{Introduction} \label{sec:intro}

It is generally thought that at the center of nearly every galaxy lies a supermassive black hole. When matter accretes onto this central black hole, an accretion disk is formed, and the galaxy becomes an active galactic nucleus (AGN; \citealt{Roger}, and references therein). Powerful jets can be launched from the AGN ($\sim10\%$ of all AGN have jets; i.e., radio-loud AGN) that can span distances from pc to Mpc and are observed across the electromagnetic spectrum. Blazars are a subclass of radio-loud AGN with relativistic jets aimed closely along our line of sight. Blazars exhibit a double-peaked, non-thermal spectral energy distribution (SED) with a lower energy peak dominated by synchrotron processes, and a higher energy peak driven by inverse-Compton  processes (IC). Blazars are classified into two categories: BL\,Lac objects, and Flat Spectrum Radio Quasars (FSRQs). BL\,Lacs have a characteristic featureless optical spectrum and are core-dominated, whereas FSRQs display emission lines and are more lobe-dominated (\citealt{Urry_1995}). Blazars are known to be highly variable across the electromagnetic spectrum, with observed variations as short as a few minutes (e.g., \citealt{Pandey_2022}; \citealt{Abeysekara_2018}). 

BL\,Lac objects are characterized along a continuum known as the blazar sequence: low-frequency-peaked BL\,Lac objects (LBLs), intermediate-frequency-peaked BL\,Lac objects (IBLs), and high-frequency-peaked BL\,Lac objects (HBLs) (\citealt{Fossati_1998}). These classification boundaries are loosely defined by the location of the synchrotron peak: $\nu_{peak} \approx 10^{13-14}$ Hz for LBLs, $\nu_{peak} \approx 10^{15-16}$ Hz for IBLs, and $\nu_{peak} \approx 10^{17-18}$ Hz for HBLs (\citealt{Nieppola_2005}). While synchrotron processes are believed to drive the observed lower energy peak of blazar SEDs, it is unclear what emission mechanism is driving the higher energy peak, and in fact, the mechanism may differ depending on where the blazar lies on the blazar sequence. Additionally, this mechanism may change depending on the activity state of the blazar. Modeling studies have shown that LBLs favor external-Compton processes (EC; e.g., \citealt{refId0}; \citealt{Sikora_2009}) and HBLs favor synchrotron-self-Compton processes (SSC; e.g., \citealt{B_ttcher_2007}; \citealt{Paggi_2009}). IBLs, however, require further study to make a conclusive statement (e.g., \citealt{B_ttcher_2013}; \citealt{refId0}). Therefore, observing and modeling IBLs during both flaring and non-flaring periods is crucial to understanding the mechanisms responsible for the observed high-energy emission originating from these objects.

BL\,Lacertae, the prototype for the BL\,Lac categorization of blazars, is a well-studied IBL with a redshift of z = 0.069 (\citealt{1978ApJ...219L..85M}). Multiple flares have been observed from BL\,Lacertae covering the entire electromagnetic spectrum. In 2011, a rapid very-high-energy (VHE; E$>$100 GeV) flare was observed from BL\,Lacertae with the Very Energetic Radiation Imaging Array System (VERITAS; \citealt{Arlen_2012}). This prompted a multiwavelength campaign that showed elevated activity in all wavebands. In 2015, another very rapid TeV flare was observed from BL\,Lacertae with the Major Atmospheric Gamma-ray Imaging Chenrenkov (MAGIC) observatory (\citealt{2019}). This again prompted a multiwavelength study, which showed activity in all wavebands. These studies, as well as others (e.g., \citealt{Abeysekara_2018}), have shown rapid variability (minutes to day timescales) in the VHE regime. The timescales observed in these studies can constrain the physical mechanisms driving the observed emission. In particular, rapid variability places constraints on the size of the emission region, implying an extremely compact emission region (refer to Equation 5 in \citealt{Abeysekara_2018}). Although these studies have led to a better understanding of IBLs, the dominant mechanism responsible for the observed VHE emission is still under investigation. 

In this study, we present the first long-duration ($\sim$\,40 days) VHE flare observed from BL\,Lacertae. The duration and broadband spectral properties of this VHE flare challenge the one-zone SSC model and suggest that an additional external Compton component is needed to explain the observed broadband emission. 

\section{Observations} \label{sec:observe}

On October 15, 2022 (59867 MJD), \textit{Fermi}-LAT detected elevated activity from BL Lacertae (\citealt{2022ATel15688....1L}), which triggered a multiwavelength campaign. Subsequently, on October 22, 2022, the X-ray Telescope (XRT) on board the Neil Gehrels Swift Observatory detected flaring activity from BL\,Lacertae (\citealt{2022ATel15730....1P}). This elevated X-ray activity was coincident with optical brightening of the source as observed by the Whole Earth Blazar Telescope Collaboration (\citealt{2022ATel15684....1M}) and the 0.4 m telescope network of Las Cumbres Observatory (\citealt{2022ATel15725....1A}). 

We are aware of other follow-up observations of this event. For example, multiple radio observatories, including the 32m Medicina Radio Telescope (\citealt{2022ATel15705....1M}) and the RATAN-600 radio observatory (\citealt{2022ATel15709....1T}), performed follow-up observations of BL\,Lacertae. These radio observations did not show significant brightening of the source. However, the measured radio flux was still very high compared to the radio flux of BL\,Lacertae between 2015 to 2020, which suggests that this flare may be part of long-lasting activity that started in 2021 (\citealt{2021ATel14330....1C}; \citealt{2021ATel14583....1L}; \citealt{2021ATel14777....1B}). In this study, we report VERITAS, \textit{Fermi}-LAT, \textit{Swift}-XRT, \textit{Swift}-UVOT, \textit{NuSTAR}, and select optical and radio observations of BL\,Lacerate during this flaring event. For the purposes of this work, we analyzed data between September 1, 2022 UTC (59823 MJD) and December 1, 2022 UTC (59914 MJD) inclusively, which is referred to as the ``epoch of interest" in this work. 

\subsection{VERITAS} \label{sec:VER}

VERITAS is a ground-based gamma-ray observatory located at the Fred Lawrence Whipple Observatory (FLWO) in southern Arizona, USA ($30^{\circ}$ 40' N, $110^{\circ}$ 57' W, 1.3 km above sea level; \citealt{vts_paper}). VERITAS is an array of four 12-meter imaging atmospheric Cherenkov telescopes (IACTs) sensitive to gamma rays with energies between $\sim$\,100\,GeV and 30\,TeV. Each telescope is equipped with a 499 photomultiplier tube (PMT) camera and can achieve an angular resolution of $\sim$\,0.1$^{\circ}$ (68\% containment at 1~TeV) and an energy resolution of $ \Delta E/E\sim$15\% (at 1~TeV) (\citealt{Holder_2006}; \citealt{christiansen2017characterization}). VERITAS is  capable of detecting ($>5\sigma$) an astrophysical source with a flux of 1\% that of the Crab Nebula in $<$25 hours of exposure (\citealt{Park15}). 

During the epoch of interest, the initial VERITAS observations of BL\,Lacertae were taken starting September 25, 2022 UTC (MJD 59847) as part of an ongoing IBL monitoring program. In response to the elevated GeV activity observed by \textit{Fermi}-LAT on October 15, 2022 (\citealt{2022ATel15688....1L}), VERITAS conducted follow-up observations in addition to the monitoring effort. After data-quality selection (good weather conditions, zenith angle $<35^{\circ}$, all telescopes operational, no moonlight conditions, no hardware issues), the source was observed for a total of $\sim$\,9.8\,hr during the epoch of interest after correcting for dead time. VERITAS does not typically observe during bright moon periods, which is reflected by gaps in the observations. The observations were conducted via a standard VERITAS ``wobble" observing mode with a $0.5^{\circ}$ offset (\citealt{FOMIN1994151}). These data were analyzed using the VEGAS analysis package \citep{VEGAS} using a predetermined set of cuts optimized for low energy sensitivity (see e.g., \citealt{Archambault_2014}) given that BL\,Lacertae is historically a soft source ($\Gamma\sim3.3$) in the VHE regime (e.g.,  \citealt{Abeysekara_2018}). These VEGAS results were then independently confirmed with the Eventdisplay analysis package \citep{Maier17}.

\begin{figure*}[h]
    \centering
    \includegraphics[width = \linewidth]{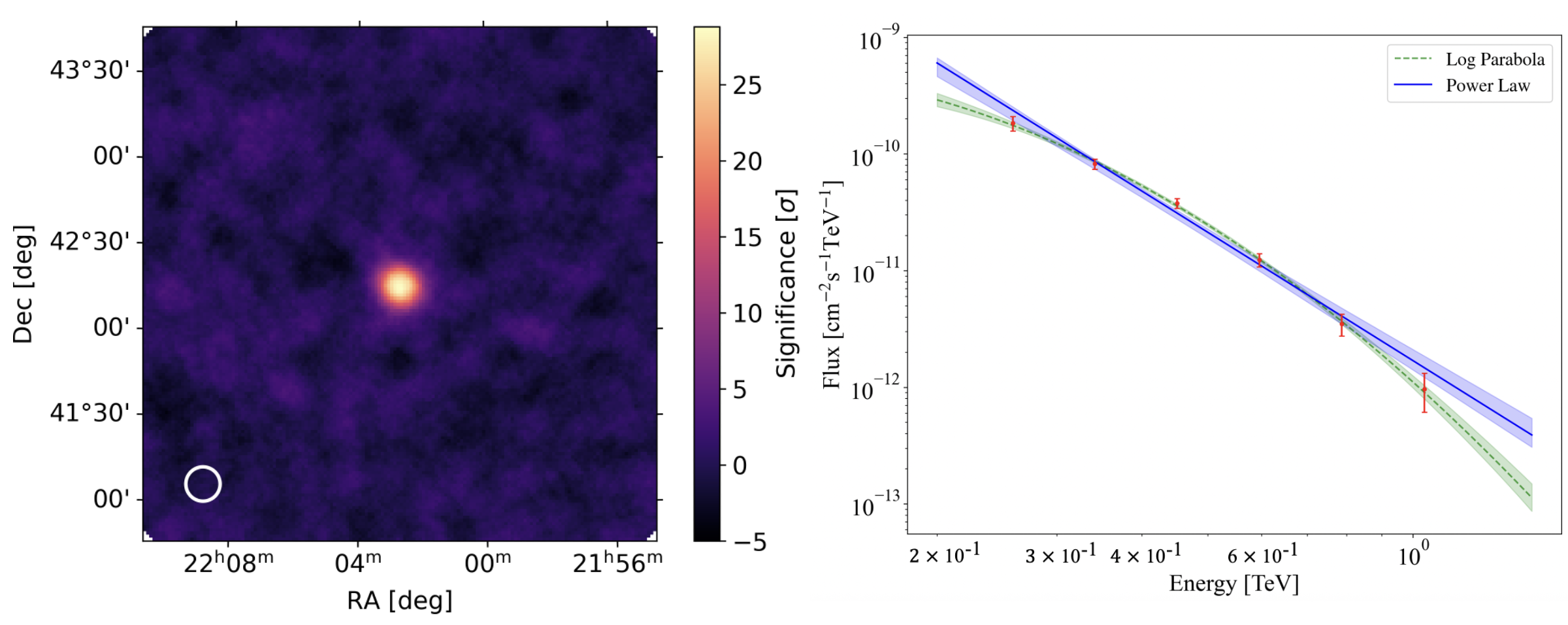}
    \caption{\textit{Left:} Skymap of the significance observed by VERITAS during the epoch of interest (September 1, 2022 UTC - December 1, 2022 UTC; 59823 - 59914 MJD inclusive) from the direction of BL\,Lacertae. The extent of the VHE source is consistent with the VERITAS point-spread function (PSF). The PSF in this analysis is reduced from prior publications due to the use of the Image Template Method $\gamma$-ray reconstruction (\citealt{christiansen2017characterization}). The PSF reference size is indicated by the white ring located in the bottom left corner of the map. \textit{Right:} The VHE photon spectrum of BL\,Lacertae observed by VERITAS (red points). The best-fit power-law function is indicated by the solid blue line, and shaded blue region corresponding to a $1\sigma$ bounds. The best-fit log-parabola function is indicated by the dashed green line, and shaded green region corresponding to a $1\sigma$ bounds.}
    \label{fig:VERITAS}
\end{figure*}

Overall, VERITAS measured 620\,$\pm$\,30 $\gamma$-ray-like events from the source region, with $N_{ON}=887$ ON events, $N_{OFF}=2941$ OFF events, and an off-source normalization factor ($\alpha$) of 0.091. This corresponds to a statistical significance of 28 standard deviations ($\sigma$) above the background using the prescription described in \cite{1983ApJ...272..317L} (refer to equation 17 therein). The significance map for these VERITAS observations is shown in Figure ~\ref{fig:VERITAS}, with the map centered on the position of BL\,Lacertae. The time-average photon spectrum is shown in Figure ~\ref{fig:VERITAS}. This spectrum is fit with a power law function of the form $dN/dE = I_0 (E/E_{0})^{-\Gamma}$ where $E_{0}$ = 0.433 TeV is set equal to the decorrelation energy of the spectrum. The best-fit flux normalization was found to be $I_0 = (3.4 \pm 0.5_{\mathrm{stat}}) \times 10^{-11}$ cm$^{-2}$ s$^{-1}$ TeV$^{-1}$ with a spectral index of $\Gamma = 3.6 \pm 0.1_{\mathrm{stat}}$. The average integral flux above a 350 GeV threshold during the epoch of interest was $(1.1 \pm 0.1) \times 10^{-11}$ cm$^{-2}$ s$^{-1}$, or $\sim$10\% of the Crab Nebula flux above the same energy threshold. The $\chi^2 / \mathrm{n.d.f}$ = 10.2/4 = 2.6 suggests that this power-law fit suboptimally describes the shape of the spectrum. Therefore, this spectrum is also fit with a log parabola function of the form $dN/dE = N_0 (E/E_{0})^{-\alpha - \beta \text{ln}(E/E_{0})}$. The best-fit normalization was found to be $N_0 = (3.7 \pm 0.1_{\mathrm{stat}}) \times 10^{-11}$ cm$^{-2}$ s$^{-1}$ TeV$^{-1}$, with a spectral index of $\alpha = 3.5 \pm 0.1_{\mathrm{stat}}$ and a curvature of $\beta = 1.1 \pm 0.1_{\mathrm{stat}}$. This fit results in a $\chi^2 / \mathrm{n.d.f}$ = 0.7/3 = 0.2. The F-test suggests that the log-parabola fit is weakly preferred over the power law fit (p = 0.0085). Refer to section \ref{sec:modeling} for further spectral fitting and modeling discussion.

A lightcurve of the observed VHE flux above 350 GeV, binned in daily intervals, is shown in Figure ~\ref{fig:mwl_lc}. During the epoch of interest, three days of observations (MJD 59847, 59870, 59877) did not yield a detection. We calculated the 99\% confidence level upper limits on the gamma-ray flux of the source during these periods using the Helene method (\citealt{HELENE1983319}; \citealt{1984NIMPA.228..120H}). To test for intra-night variability, individual nights were binned by observation run (30 minutes). Each night that had multiple observation runs were then fit with a constant function. The results of these fits suggest that a constant flux is consistent with the observed flux each night given the statistical uncertainties, and therefore, we do not detect significant intra-night variability.

\newpage

\subsection{Fermi-LAT}

The \emph{Fermi}-Large Area Telescope (LAT; \citealt{Fermi_LAT}) is a pair-conversion telescope sensitive to gamma-ray photons in the energy range from 20~MeV to above 500 GeV. Primarily operating in survey mode, it scans the entire sky every three hours. In this work, \textit{Fermi}-LAT photons detected during the epoch of interest are analyzed. Throughout the analysis, the \textit{Fermi} Science Tools version $11-05-03$,\footnote{\url{http://fermi.gsfc.nasa.gov/ssc/data/analysis/software} (accessed on 02/27/2023)}  and \textit{FERMIPY} version 1.0.1~\footnote{\url{http://fermipy.readthedocs.io} (accessed on 02/27/2023)} \citep{wood2017fermipy} are used, in conjunction with the latest \textit{PASS} 8 IRFs~\citep{atwood2013pass}. 

Photons with energies between 100 MeV and 100 GeV detected within a region of interest (RoI) of radius 15$^{\circ}$ centered on the location of BL\,Lacertae are selected for analysis. Furthermore, only photon events within a maximum zenith angle of 90$^{\circ}$ are selected in order to reduce contamination from background photons from the Earth's limb, produced from the interaction of cosmic-rays with the upper atmosphere.
The contributions from the isotropic and Galactic diffuse backgrounds are modeled using the most recent templates for isotropic and Galactic diffuse emission, iso\_P8R3\_SOURCE\_V2\_v1.txt and gll\_iem\_v07.fits respectively.

Sources in the 4FGL-DR3 catalog within a radius of $20^{\circ}$ from the source position of BL\,Lacertae are included in the model with their spectral parameters fixed to their catalog values.
This accounts for the gamma-ray emission from sources lying outside the RoI which might yet contribute photons to the data, especially at low energies, due to the size of the \textit{Fermi}-LAT point spread function. The normalization factor for both the isotropic and Galactic diffuse emission templates are left free along with the spectral normalization of all modeled sources within the RoI. Moreover, the spectral shape parameters of all modeled sources within 3$^{\circ}$ of BL\,Lacertae are left free to vary while those of the remaining sources were fixed to the values reported in the 4FGL-DR3 catalog \citep{4fgl_dr3}. A binned likelihood analysis is performed in order to obtain the spectral parameters best describing the model during the period of observation, using a spatial binning of 0.1$^{\circ}$ per pixel and 8 energy bins per decade.

The source is detected at a high statistical significance of more than 167$\sigma$ (TS = 27895) during the entire epoch of interest. The average flux between 0.1\,-\,300 GeV during the GeV flare epoch of interest was $(6.9 \pm 0.5) \times 10^{-6}$ cm$^{-2}$ s$^{-1}$. This corresponds to a flux increase of more than a factor of 15 relative to the average flux reported in the 4FGL-DR3 catalog (\citealt{4fgl_dr3}). The spectrum was found to be best modeled by a log parabola of the form $dN/dE = N_0 (E/E_{0})^{-\alpha - \beta \text{ln}(E/E_{0})}$. The best-fit spectral parameters obtained for BL\,Lacertae are $N_{0} = (3.40 \pm 0.05) \times 10^{-10}$ $\text{cm}^{-2}\text{s}^{-1} \text{MeV}^{-1}$, $\alpha = 1.96 \pm 0.01$ and $\beta = 0.05 \pm 0.01$. For comparison, the 4FGL-DR3 catalog spectral values for this source are $N_{0} = (5.28 \pm 0.05) \times 10^{-11}$ $\text{cm}^{-2}\text{s}^{-1} \text{MeV}^{-1}$, $\alpha = 2.11 \pm 0.01$ and $\beta = 0.59 \pm 0.04$. 

\subsection{X-ray Facilities}

\textit{Swift-XRT} --- The X-ray Telescope (XRT) on board the Neil Gehrels Swift Observatory is sensitive to photons with energies between 0.2 - 10\,keV (\citealt{2004AIPC..727..637G}; \citealt{Burrows_2005}). \textit{Swift}-XRT observed BL\,Lacertae as part of an ongoing monitoring program, but increased its observation cadence in response to the elevated GeV activity observed by \textit{Fermi}-LAT from the source on October 15, 2022 (\citealt{2022ATel15688....1L}). In late October 2022, \textit{Swift}-XRT detected flaring activity from BL\,Lacertae (\citealt{2022ATel15730....1P}). Target of Opportunity (ToO) observations were conducted following the detection of this increased X-ray activity and the initial GeV flare detected by \textit{Fermi}-LAT. 

In total, \textit{Swift}-XRT took 26 observations of BL\,Lacertae during the epoch of interest. The November 11, 2022 (OBSID 0001492502) observation was excluded from this analysis due to an unforeseen interruption during observations, resulting in only a 96 second exposure. The remaining 25 observations resulted in a cumulative exposure of 32.2 ks. These observations were conducted in photon-counting (PC) mode, and were analyzed using the \texttt{HEASoft} package (v6.32.1\footnote{\url{https://heasarc.gsfc.nasa.gov/docs/software/heasoft/}}). Given that BL\,Lacertae is relatively close to the galactic plane (with a Galactic latitude $b = -10.44^{\circ}$), galactic absorption of neutral hydrogen is corrected using a Galactic column density of $N_{H}~= ~ 2.92\times 10^{21}$ cm$^{-2}$ (\citealt{Willingale_2013}). 

A spectrum was then constructed and fit with \texttt{xspec} with an absorbed power law. A good fit (W-stat$/ \mathrm{ndf}$~=797/728) is found for a photon index of $2.0 \pm 0.1$. The time-averaged unabsorbed flux from the source in the 0.3--10 keV range for these observations is $5.13 \pm~0.07$~$\times 10^{-11}$ erg cm$^{-2}$ s$^{-1}$. A lightcurve of the absorption-corrected flux values is shown in Figure~\ref{fig:mwl_lc}. These data are also used in the modeling of the broadband SED of the source in Section~\ref{sec:modeling}. The results of this analysis were independently confirmed using the \textit{Swift}-XRT Online Tools\footnote{\url{https://www.swift.ac.uk/user_objects/docs.php}} (\citealt{Evans_2009}; \citealt{Evans_2007}) which utilized the \texttt{HEASoft} v.6.32 pipeline. 

\textit{NuSTAR} --- The Nuclear Spectroscopic Telescope Array (\textit{NuSTAR}) is a satellite-based focusing high-energy X-ray telescope sensitive to photons with energies between 3 and 79\,keV (\citealt{Harrison_2013}). \textit{NuSTAR} observed BL\,Lacertae once during the epoch of interest on November 28, 2022 (59911 MJD, Obs ID 90801633002) yielding exposures of 20.8\,ks and 20.6\,ks with the instrument Focal Plane Modules A (FPMA) and FPMB, respectively. Data were reduced for both modules using the \textit{NuSTAR} Data Analysis Software package (\texttt{NuSTARDAS}\footnote{\url{https://heasarc.gsfc.nasa.gov/docs/nustar/analysis/nustar_swguide.pdf}}) within \texttt{HEASOFT} v6.32.1 package. To produce clean and calibrated event files, we used the \texttt{nupipeline} tools prescription. We then extracted source and background spectra for each module using the \texttt{nuproducts} pipeline and the following regions: a circular source extraction region of 30 pixel  ($\sim$70") radius centered on the coordinates of the target and a background extraction region of 30 pixel radius positioned away from the target in the same frame. The spectra were rebinned with a minimum of 20 counts per bin. The count rates for FPMA and FPMB are 0.673$\pm$0.006 counts s$^{-1}$ and 0.631$\pm$0.006 counts s$^{-1}$ respectively. These were jointly fit to an absorbed power law with \texttt{XSPEC} v12.13.1. Similar to the \textit{Swift}-XRT analysis, the Galactic column density was fixed to $N_{H}~= ~ 2.92\times 10^{21}$ cm$^{-2}$ (\citealt{Willingale_2013}).

\subsection{Optical and Ultraviolet Facilities}

\textit{Swift-UVOT} --- The Ultraviolet/Optical Telescope (\textit{Swift}-UVOT) on board the Neil Gehrels Swift Observatory is a photon-counting telescope sensitive to photons with energies ranging from 1.9 to 7.3\,eV \citep{2005SS_Roming}. The observations were performed in parallel to the XRT measurements. Similar to the \textit{Swift}-XRT analysis, we excluded the November 11, 2022 (OBSID 0001492502) observations due to interruptions during the observing run. \emph{Swift}-UVOT observations in all six filters (UW1, UW2, UM2, U, B, and V) were analyzed using the \texttt{HEASoft v.6.32 uvotsource} routine by selecting source photons from a circular region of radius 5'' centered on BL\,Lacertae. The background was estimated by selecting a 20''~radius circular region away from the blazar that contains no obvious sources in any band. The measured optical-UV magnitudes were then converted to fluxes using the zero-points given by \citet{Poole_2007} and are shown in Figure ~\ref{fig:mwl_lc}. These are corrected for Galactic extinction using the approach of \citet{McCall2007} assuming a reddening found in \citet{Schlafly_2011}.

\textit{FLWO} ---  BL\,Lacertae was observed during the epoch of interest by the 48” optical telescope at FLWO as part of a long-standing study of approximately 70 blazars. Data were collected using SDSS r', SDSS i', Harris V, and Harris B filters. During the epoch of interest, a total of 102 photometric images of BL\,Lacertae were collected over 26 nights, of which 92 pass quality control tests. 

Aperture photometry with an extraction radius of 3.38 arcseconds (5 pixels) is used to measure source magnitudes calibrated against stars from the AAVSO Photometry All-Sky Survey (APASS) catalog (\citealt{Henden2016VizieROD}) in each image and checked against a reference star. For BL\,Lacertae, the estimated systematic errors are $0.4, 0.4, 0.4, 0.3$ mag for the SDSS r', SDSS i', Harris V, and Harris B filters, respectively. The flux is determined at the central wavelength of each filter. These are then corrected for Galactic extinction assuming a reddening found in \citealt{Schlafly_2011}. These observations can be seen in Figure \ref{fig:mwl_lc}. Overall, we observe a gradual optical brightening of BL\,Lacertae during the epoch of interest.

\textit{ATLAS} --- Optical observations of BL\,Lacertae were made with the Asteroid Terrestrial-impact Last Alert System (ATLAS; \citealt{ATLAS_main}) during the epoch of interest. ATLAS is a high-cadence all-sky survey system consisting of four 0.5-meter telescopes that scan the entire sky periodically (typical observations set on a two-day cadence). These telescopes are located in Hawaii (Mauna Loa, Hawaii and Haleakala, Maui), South Africa (Sutherland Observing Station), and Chile (El Sauce Observatory, Rio Hurtado). ATLAS is optimized to be an efficient system for finding variable and transient sources, including potentially dangerous asteroids.  

For the purpose of this study, ATLAS observations of BL\,Lacertae were taken in the R-band (transmission curve centered at 679 nm; see Tables 1 and 2 of \citealt{ATLAS_main}). A forced photometry system is used for data processing purposes, and is described in \cite{ATLAS_main} and \cite{ATLAS_smith}. The  results of these observations can be seen in Figure ~\ref{fig:mwl_lc}. Similarly to the FLWO optical observations, we observe a gradual optical brightening of BL\,Lacertae during the epoch of interest.

\newpage
\subsection{Radio Facilities}

\textit{VLBA} --- We have analyzed Very Long Baseline Array (VLBA) data at 43~GHz obtained for BL\,Lacertae within the Blazar Entering the Astrophysical Multi-Messenger Era program (BEAM-ME~\footnote{www.bu.edu/blazars/VLBAproject.html}). The VLBA observed BL\,Lacertae three times during the epoch of interest. The data reduction was performed using the Astronomical Image Process System (AIPS) and {\it Difmap} software packages as described in \citealt{Jorstad_2017}. 

\begin{figure*}[p]
    \centering
    \includegraphics[width = \linewidth]{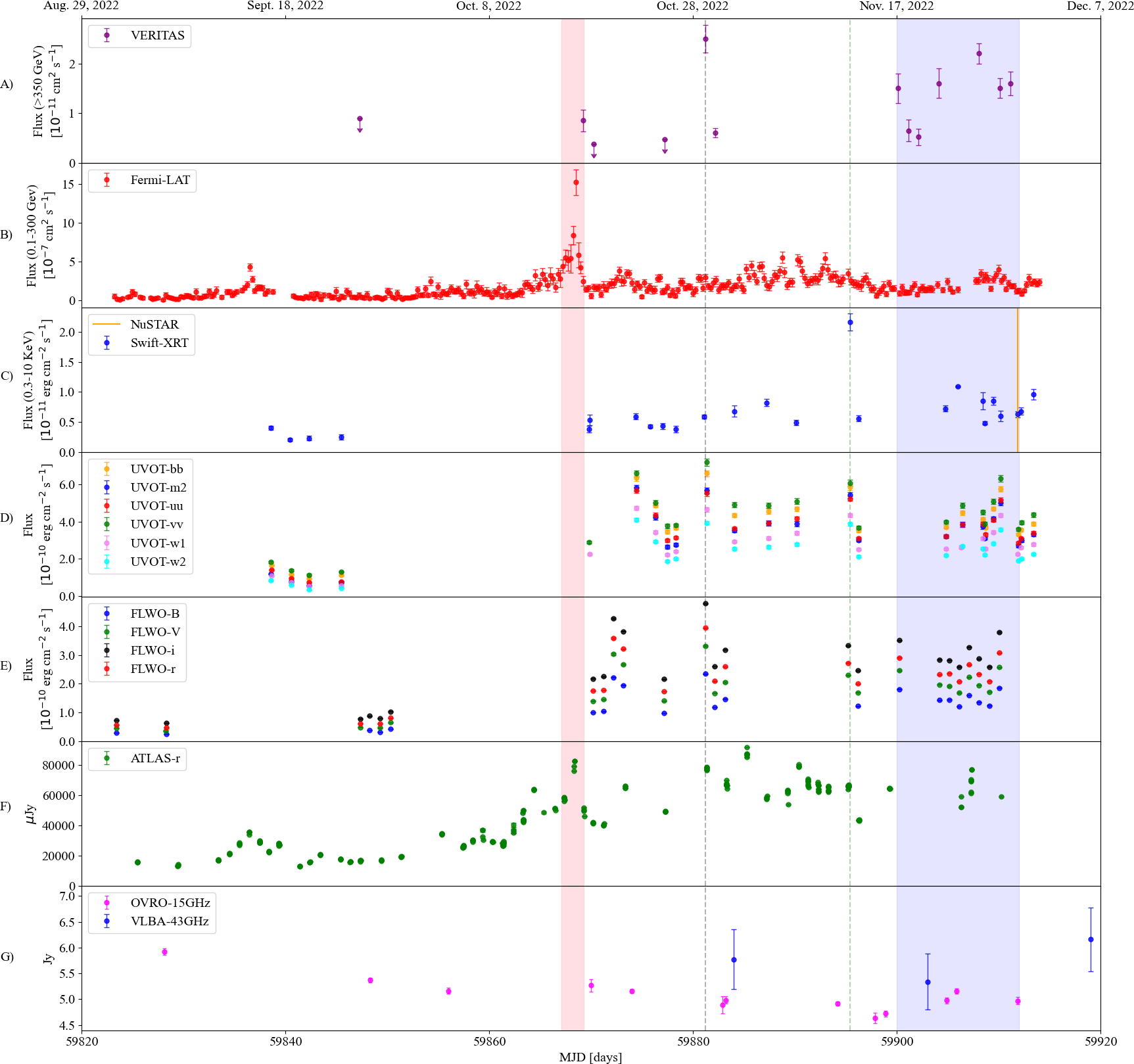}
    \caption{Multiwavelength observations of BL\,Lacertae between MJD~59823 and MJD~59914 (September 1, 2022 - December 1, 2022). The pink and purple narrow bands correspond to the GeV flare epoch and the elevated VHE activity epoch respectively. Vertical dashed grey line indicates October 29, 2022, associated with short-term elevated VHE activity. Vertical dashed green line indicated November 12, 2022, associated with enhanced X-ray activity. \textbf{(A)} VHE lightcurve for VERITAS observations above an energy threshold of 350 GeV binned per day. Upper limits at 99\% confidence level are quoted for observations with a significance lower than $5\,\sigma$. \textbf{(B)} \emph{Fermi}-LAT lightcurve in 6 hour bins in the 0.1--300 GeV band. Gaps correspond to a significance less than $2\sigma$. \textbf{(C)} \emph{Swift}-XRT absorption-corrected flux observations in the 0.3--10 keV band. The timing of the NuSTAR observation is indicated by the orange vertical line. \textbf{(D)} \emph{Swift}-UVOT flux observations in the w1, w2, m2, uu, bb, and vv band filters. \textbf{(E)} Optical lightcurve in r, i, V, and B filters using data from the 48” optical telescope at the FLWO. The errors plotted are statistical uncertainties only. \textbf{(F)} Optical flux observations with ATLAS in the r-band. \textbf{(G)} Radio flux observations at 15 GHz with OVRO and 43 GHz with VLBA.}
    \label{fig:mwl_lc}
\end{figure*}

We have calculated the total flux density, $S_{43GHz}$, degree of polarization, $P_{43GHz}$, and position angle of polarization, $\chi_{43GHz}$, for each observation by integrating the flux density over images in Stokes I, Q, and U parameters. During this period there is no significant flux density or degree of polarization variability at 43 GHz, with the average value of $S_{43GHz}\sim$6~Jy and $P_{43GHz}\sim$5.5\%. The position angle of polarization changes from 3$^o$ to 22$^o$, although all values of $\chi_{43GHz}$ are close to the jet direction $\Theta_{jet}$=$-$170.8$\pm$7.2$^o$ (\citealt{Weaver_2022}). This supports the idea of a toroidal structure of the magnetic field in the BL\,Lacertae jet (\citealt{G_mez_2016}). These measurements are plotted in Figure \ref{fig:mwl_lc}.

\textit{OVRO} --- The Owens Valley Radio Observatory (OVRO) is located in Owens Valley, California, USA, and hosts a 40-m radio dish antenna with off-axis dual-beam optics and a cryogenic receiver with 2~GHz equivalent noise bandwidth centered at 15~GHz. Atmospheric and ground contributions as well as gain fluctuations are removed with the double switching technique \citep{1989ApJ...346..566R}. Observations are conducted in an ON-ON fashion so that one of the beams is always pointed on the source. Since May 2014, a new pseudo-correlation receiver is used with a 180~degree phase switch. Relative calibration is achieved with a temperature-stable noise diode to compensate for gain drifts. 3C~286 is the primary flux density calibrator with an assumed value of 3.44~Jy \citep{1977A&A....61...99B}, and DR21 is used as secondary calibrator source. \citet{2011ApJS..194...29R} describes details of the observation and data reduction schemes.

BL\,Lacertae is routinely observed as part of the OVRO 40-m blazar monitoring program with an average cadence of two times per-week (\citealt{2011ApJS..194...29R}). Thirteen observations fall within the epoch of interest for this study. The lightcurve for BL\,Lacertae, shown in Figure \ref{fig:mwl_lc}, indicates a gradual dimming of the source at radio frequencies, with no obvious elevated activity during the epoch of interest.

\section{Multi-wavelength Analysis}

\subsection{Multiwavelength Lightcurve} \label{lc}

Compiling all of the aforementioned observations, a multiwavelength lightcurve was built and is shown in Figure ~\ref{fig:mwl_lc}. The lightcurve covers the time interval between MJD~59823 and MJD~59914. This corresponds to September~1, 2022 until December~1, 2022 -- the epoch of interest.

\subsubsection{GeV Flare Epoch} \label{sec:GeV}

As described in \citet{2022ATel15688....1L}, on October 15, 2022 UTC, increased GeV activity was detected in the direction of BL\,Lacertae by \textit{Fermi}-LAT. A Bayesian block analysis (\citealt{2013ApJ...764..167S}) was used on the \textit{Fermi}-LAT lightcurve to determine the beginning and the end of this elevated GeV state. This was performed using the Astropy Python Package (\citealt{astro}; \citealt{Price-Whelan_2018}) with an empirical prior equivalent to a false alarm probability $p_0 = 3\times10^{-7}$ ($5\sigma$ confidence level). The edges (change points) for this elevated GeV state were found to be October 15, 2022 UTC (MJD 59867) and October 17, 2022 UTC (MJD 59869) rounded to the whole day. For the purposes of this study, we will refer to this as the ``GeV flare epoch" and it is represented by a pink shaded region in Figure~\ref{fig:mwl_lc}. Unfortunately, most multiwavelength observations did not have coverage during this GeV flare epoch, with only VERITAS catching the end of the epoch and ATLAS observing around the peak of the GeV flare. 

\begin{figure*}[h]
    \centering
    \includegraphics[width = 5in]{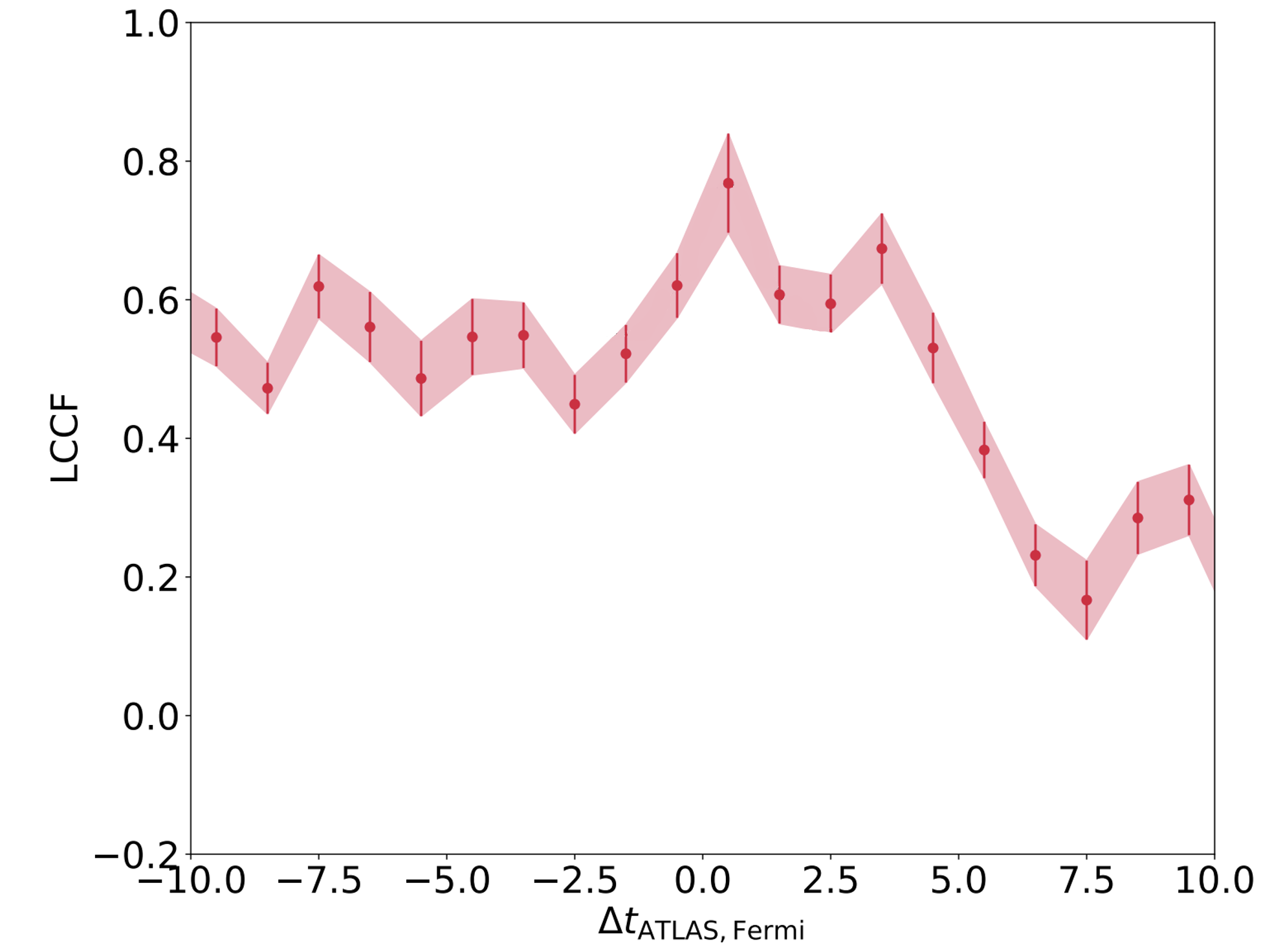}
    \caption{Local cross-correlation function (LCCF, \citealt{Welsh_1999}) calculated between the observed GeV emission and ATLAS measurements during the GeV flare epoch. The x-axis represents the time delay, $\Delta t _{ATLAS,Fermi} = t_{ATLAS} -t_{Fermi}$, in days between the \textit{Fermi}-LAT and ATLAS lightcurves (\citealt{2023Galax..11...81A}). The corresponding shaded regions indicate the error bounds of the LCCFs. There is very strong correlation between them, noting the most significant of these peaks at a correlation consistent with zero delay. This could suggest a co-spatial production site for the observed optical and GeV emission.}
    \label{fig:Correlation}
\end{figure*}

From visual inspection, potential correlation is noted between the GeV observations and ATLAS measurements. A Local Cross Correlation Function (LCCF, \citealt{Welsh_1999}) is applied to the \textit{Fermi}-LAT and ATLAS lightcurves. As can be seen in Figure~\ref{fig:Correlation}, there is very strong correlation between them, noting the most significant of these peaks at a correlation consistent with zero delay. This could suggest a co-spatial production site for the observed optical and GeV emission. This behavior can be explained by single-zone SSC processes, where an injection of a single population of relativistic particles synchrotron self-Compton scatter off of optical photons to produce co-spatial GeV photons. Unfortunately, due to the lack of multi-band coverage during this epoch, the single-zone SSC model hypothesis cannot be fully tested nor efficiently constrained through SED modeling efforts.

\subsubsection{Long-Duration Elevated VHE Activity Epoch} \label{sec:dur}

Although the observed GeV flare was short lived ($\sim$\,2 days), VERITAS also detected elevated VHE activity from BL\,Lacertae up to $\sim$\,40 days after the GeV flare. We define ``elevated VHE activity" for BL\,Lacertae if we detect the source within a nights' worth of observations (ranging from 30 minutes to 2 hrs exposure). Typically, BL\,Lacertae, as well as other LBLs and IBLs, requires long exposures ($\gg10$\,hrs) to be detected ($>5\sigma$) in the VHE regime during its low state. It is not until the source is in an elevated VHE state that it can be detected within an observation run (usually 30 minutes). Therefore, we define an ``elevated VHE activity epoch" to be the beginning of continuous nightly detections ($\le 3$ nights between detections) of the source (November 17, 2022 UTC) until the end of these continuous nightly detections (November 28, 2022) inclusively. Note also that VERITAS detected the source on October 29, 2022 UTC and October 30, 2022 UTC (see Section \ref{sec:VERFLARE}). These observations are not included in the elevated VHE activity epoch as we are unable to determine the source's activity from 2022 October 31 to 2022 November 17. Large gaps exist in these data due to moonlight (as VERITAS does not typically observe during bright moon periods) and poor weather. Observations of the source ended on 2022 November 28 as BL\,Lacertae became unobservable at modest zenith angles ($\theta_{z} < 55^{\circ}$). 

This ``elevated VHE activity epoch" is therefore defined to be from November 17, 2022 (59900 MJD) to November 28, 2022 UTC (59911 MJD) inclusive, and is represented by the purple shaded region in Figure~\ref{fig:mwl_lc}. This is the longest sustained elevated VHE activity epoch ever detected from BL\,Lacertae as all other VHE outbursts from the source have been detected on rapid timescales ($\sim$\,minutes to a day; refer to Section~\ref{sec:intro} for examples). Additionally, the end of this epoch corresponds to $\sim$\,40 days after the initial elevated GeV activity epoch ended. Most previously observed VHE flares from BL\,Lacertae correspond temporally with GeV activity (ex. \citealt{Abeysekara_2018}), but this is not the case for this VHE flare. It is unclear what physical mechanisms power this long-term VHE activity while the observed GeV activity decreases. These observations could potentially be explained by a new high-energy particle population being injected into the emission region, hardening the electron distribution. Fortunately, all observatories in this study had at least one observation during the elevated VHE activity epoch, but it should be noted these observations are not all strictly simultaneous. Therefore, this presents an opportunity to test proposed flare mechanism theories via broadband spectral modeling. Both SSC and SSC+EIC models are fit to the data (see section \ref{sec:modeling}), and are the main focus of the present work.

\subsubsection{Rapid Multiwavelength Flare Activity} \label{sec:VERFLARE}

In addition to the GeV flare epoch (see Section \ref{sec:GeV}) and the long-duration elevated VHE activity epoch (see Section \ref{sec:dur}), rapid flaring activity has been identified via visual inspection of the multiwavelength lightcurve (See Figure \ref{fig:mwl_lc}).

VERITAS detected elevated activity from the source on October 29, 2022 UTC (indicated by a grey dashed line in Figure \ref{fig:mwl_lc}) and October 30, 2022 UTC. On October 29, 2022, VERITAS observed the source for $\sim$\,1 hour, consisting of two 30-minute observation runs. An analysis of these data yields a $\sim$\,13$\sigma$ detection of the source. The average integral flux above a 350 GeV threshold was $(2.5 \pm 0.1) \times 10^{-11}$ cm$^{-2}$ s$^{-1}$, or $\sim$\,22\% of the Crab Nebula flux above the same energy threshold. This was the highest measured flux from the source during the epoch of interest. On October 30, 2022, VERITAS observed the source for $\sim$\,2 hours, consisting of four 30-minute observation runs. An analysis of these data yields a $\sim$\,8$\sigma$ detection of the source. The average integral flux above a 350 GeV threshold was $(6.0 \pm 0.1) \times 10^{-12}$ cm$^{-2}$ s$^{-1}$, or $\sim$\,5\% of the Crab Nebula flux above the same energy threshold. As discussed in Section \ref{sec:VER}, no intra-night variability is detected from the source.

During this time period (October 29 - October 30 UTC inclusive), elevated optical activity is detected via visual inspection, while no obvious signs of flaring activity are noted in the GeV and X-ray measurements. This would indicate to a more complex scenario as compared to the single-zone SSC model proposed previously to explain the multiwavelength activity during the GeV flare epoch. However, this  multiwavelength activity has previously been observed from this source during the 2011 outburst (\citealt{Arlen_2012}). During the 2011 flare, VERITAS observed a rapid TeV flare from BL\,Lacertae after strong GeV flaring activity was detected from the source $\sim$\,30 days prior by \textit{Fermi}-LAT. Similar to the observations discussed in this work, neither enhanced GeV nor X-ray activity was detected from the source during the TeV flare, with evidence of flux variations at optical and UV wavelengths. 

This multiwavelength activity could indicate a more complex jet structure with multiple SSC zones or relativistic particle populations required to explain the observed emission. Similarly discussed in \citealt{Arlen_2012}, independent of any models assumed, the short-lived behavior of this enhanced activity requires the emission region to be small given $R\leq c\tau_{d}\delta / (1+z)$, where $R$ is the radius of the emission region, $\tau_{d}$ is the decay time of the flare, $z$ is the redshift of the source, and $\delta$ is the Doppler factor of the jet. This is in contrast to the unique long-duration elevated VHE activity epoch -- the primary focus of this work. 

Lastly, elevated X-ray activity is noted on November 12, 2022 UTC, which is indicated by a green dashed line in Figure \ref{fig:mwl_lc}. With no major  multiwavelength counterparts, this enhanced X-ray activity can be best explained by an injection of a new high-energy particle population leading to a hardening of the electron distribution. Assuming SSC mechanisms, these X-ray photons act as seed photons for IC processes, which would then be reflected in the VHE regime. Unfortunately, this was during the bright moon period, so no VERITAS observations of the source were taken. Previous multiwavelength studies of BL\,Lacertae have shown variability in the X-ray regime with no associated counterparts, similar to the observations discussed in this work (\citealt{Raiteri_2009} \& \citealt{Raiteri_2010}). As discussed in \citealt{Raiteri_2010}, a complex multi-zone emission model was needed to explain the observed SED. Additionally, this study showed that the complex jet geometry can play a vital role in the spectral and flux variability that is observed from BL\,Lacertae. Therefore, the overall enhanced activity at various timescales and energies as shown in Figure \ref{fig:mwl_lc} and discussed in this section as a whole can suggest multi-zone emission models may be needed to explain the uniqueness of this flaring activity (e.g. \citealt{Abeysekara_2018} and \citealt{Hervet_2016}).

\begin{figure*}[h]
    \centering
    \includegraphics[width = \linewidth]{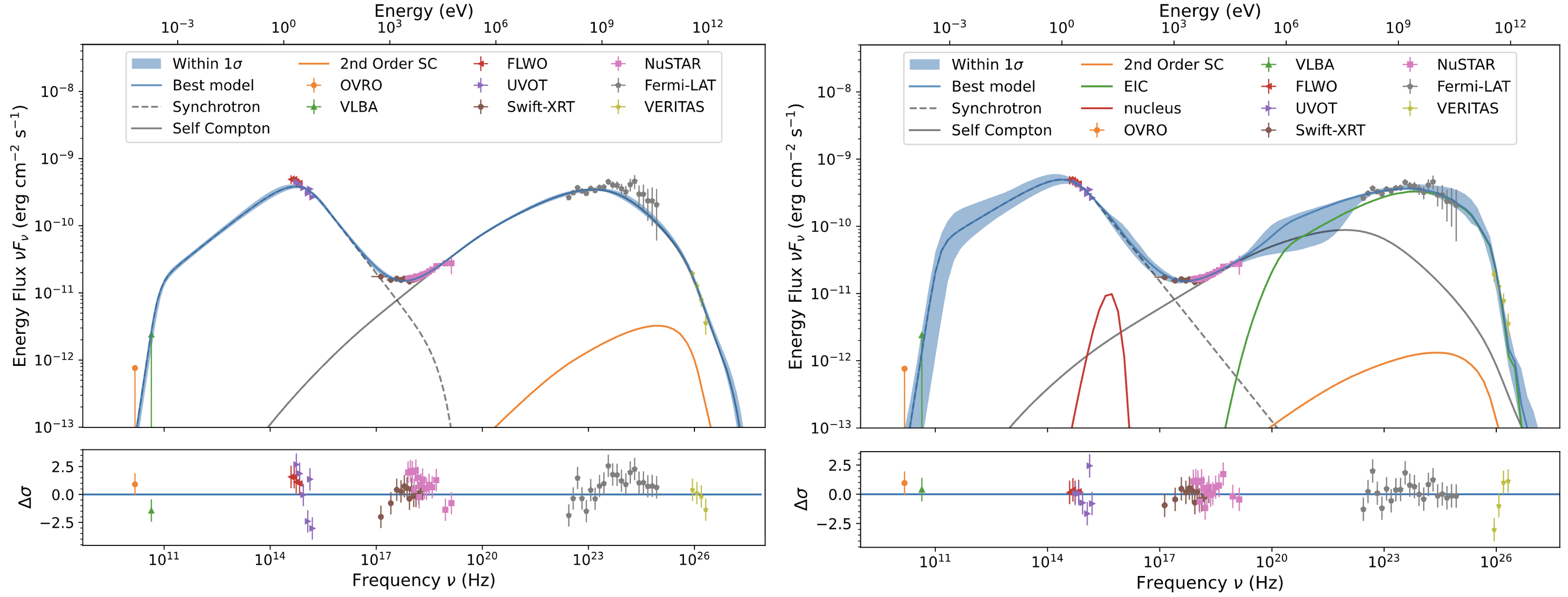}
    \caption{\textit{Left:} Broadband SED for BL\,Lacertae during the VERITAS elevated VHE activity epoch. The radio SED points from OVRO and VLBA are treated as upper limits for modeling purposes due to a larger integration region, potentially contaminated by extended jet emission. This SED is fit to a single-zone SSC model. \textit{Right:} The same SED is fit to an SSC model with an additional external inverse-Compton component.}
    \label{fig:SED}
\end{figure*}

\subsection{SED Modeling} \label{sec:modeling}

Utilizing all observations from the elevated VHE activity epoch, a broadband SED was constructed (see Figure~\ref{fig:SED}). To build the most conservative spectrum, if there were multiple observations during this epoch for a single bandpass (ie. observatories with individual bandpass filters: VLBA (43GHz), OVRO (15GHz), FLWO (B, V, i', r'), UVOT (bb, uu, vv, w1, w2, m2)), the time-averaged spectral point was calculated for that specific energy bin. In this case, both the statistical error for that spectral point and root mean square deviation (RMSD) of the flux variations within the epoch itself were calculated. For each energy bin, we then took the greater of these two uncertainties (statistical error and RMSD) and assigned this uncertainty to the spectral point to find the most conservative fit for the SED. For observatories covering multiple bandpasses (NuSTAR, \textit{Swift}-XRT, \textit{Fermi}-LAT, and VERITAS), the statistical as well as the systematic errors were considered when assigning uncertainties to their associated spectral points. The host galaxy corrections were made to the FLWO and \textit{Swift}-UVOT observations according to the pipeline ``SED Host Galaxy Correction"\citep{hervet_2024_10800919}. All other corrections to the data are described in Section~\ref{sec:observe}.

In order to interpret the multiwavelength behavior of this long-duration VHE flare from BL\,Lacertae, the broadband SED is modeled using \texttt{Bjet\_MCMC}\footnote{\url{https://github.com/Ohervet/Bjet_MCMC} (accessed on 02/21/2024)} \citep{2023_Bjet}: a public tool that can automatically fit the broadband SED of blazars. The model has previously been used to study several jetted AGN emitting in the VHE band; For example AP Librae \citep{refId0}, 1ES 1215+303 \citep{Valverde_2020}, TON~599 \citep{Adams_2022}, and OJ 287 \citep{acharyya2024multiwavelengthstudydecipher2017}. The radio SED points from OVRO and VLBA are treated as upper limits for modeling purposes due to a larger integration region, potentially contaminated by extended jet emission. Extragalactic background light (EBL) corrections are applied within \texttt{Bjet\_MCMC} itself to both the \textit{Fermi}-LAT and VERITAS spectral points according to the prescription described in \citealt{Franceschini_2017} using the redshift of $z=0.069$. 

A simple one-zone SSC radiation model is considered and consists of a single blob with a broken power law differential electron energy density (EED), a scenario  commonly used to model BL\,Lac objects. The SED fit is performed using the Markov-Chain Monte-Carlo (MCMC) method, also implemented in several modeling studies of blazars; For example \cite{Tramacere_2011} and \cite{10.1093/mnras/staa3163}. This model consists of nine parameters: the Doppler factor, $\delta$, particle density normalization, $K$, the particle spectrum indices from below and above $\gamma_{break}$, $n_1$ and $n_2$ respectively, the minimum and maximum Lorentz factors of the EED, $\gamma_{\text{min}}$ and $\gamma_{\text{max}}$ respectively, the break Lorentz factor of the electron distribution, $\gamma_{\text{break}}$, the magnetic field, $B$, and the blob radius, $R$. More details of the parameters can be found in \cite{2023_Bjet}. All the parameters are free to vary, with inherent physical constraints set within the code. The MCMC sampling was run for 6000 steps with 50 walkers and with the first 200 steps discarded as a burn-in period. As seen in Figure~\ref{fig:SED}, a simple one-zone SSC  model describes the data moderately well over the energy range considered ($\chi^2 / \mathrm{ndf}$~=~96.96/56~=~1.73). 

The best-fit values of the parameters are reported in Table~\ref{tab:table1}. Overall, the fitted magnetic field and Doppler factor remain broadly compatible with a one-zone SSC scenario, although they lie toward the extreme end of the parameter space typically inferred for BL\,Lac objects (\citealt{B_ttcher_2013}). Previously observed long-sustained VHE flares have reported large Doppler factors (ex. \citealt{Tavecchio_2010} \& \citealt{Acharyya_2025}). These high Doppler factors and magnetic fields could be needed to sustain this long-duration VHE activity. Additionally, the sharp electron break is stronger than the canonical synchrotron cooling break and is therefore unlikely to arise from pure radiative cooling alone (\citealt{B_ttcher_2013}). We instead interpret it as an effective phenomenological description of the particle distribution, potentially reflecting additional processes such as energy-dependent escape or non-standard acceleration. Therefore, taking these fit results in conjunction with the multiwavelength behavior previously discussed in section \ref{sec:VERFLARE}, the underlying physical mechanisms that drive the previously observed rapid variability could be different than those that drive long-duration flares. This would naturally lead us to test more complex models which we believe could better describe the observed behavior.

As a next step, a more complex SSC+EIC radiation model is considered. This model includes a thermal external inverse-Compton (EIC) component from the interaction of high-energy particles of the jet with the thermal ambient radiation field surrounding the nucleus due to the accretion disk emission reprocessed by the broad-line region (BLR). In addition to the nine parameters in the one-zone SSC model, this model also includes: black body temperature of the disk, $bb_{temp}$, nucleus luminosity, $L_{nuc}$, fraction of the luminosity scattered, $\tau$, and the distance of the blob to the super massive black hole (SMBH), $D_{BH}$. All of the parameters are free to vary, with inherent physical constraints set within the code, except for the nucleus luminosity in which we set an upper limit. 

The nucleus of BL\,Lacertae, as well as other BL Lac objects, is thought to be a stable source. We can place an upper limit on how luminous we believe the nucleus is by taking the lowest optical flux observed. For this, we analyzed optical data from FLWO-48 over the past 10 years and found the lowest flux point in all filters. This minimum flux is then converted into a luminosity using the observed redshift ($z=0.069$) and the luminosity distance of BL\,Lacertae ($D_L=312.9$\,Mpc, assuming $H_0=69.6$) along a continuum of temperatures. The strongest constraint on the nucleus luminosity is then the minimum of nucleus luminosity for all possible nucleus temperatures. The MCMC sampling was run for 18,000 steps with 140 walkers and with 200 steps discarded as a burn-in period. The values of the parameters are reported in Table~\ref{tab:table1}.  As seen in Figure~\ref{fig:SED}, the SSC+EIC model describes the data well with a good agreement seen between the fitted model and the data over the energy range considered ($\chi^2 / \mathrm{ndf}$~=~38.80/52~=~0.75). 

When we compare these two nested models (SSC vs. SSC+EIC) using a log-likelihood ratio test (\citealt{d543aecb-cd73-36d5-9101-f08a74f8e8c6}), we find that the SSC+EIC model is preferred over the simple one-zone SSC model with a significance of $6.9\sigma$. This significant statistical preference suggests that an additional EIC component is needed to explain the observed VHE emission from BL\,Lacertae during this elevated VHE activity epoch.

Taking the results of this SSC+EIC fit as well as the overall complexity of the multiwavelength flaring activity during the entire epoch of interest (September 1, 2022 UTC (59823 MJD) to December 1, 2022 UTC (59914 MJD) inclusive; See section \ref{lc}), this suggests that a more complex model beyond the simple one-zone SSC model is needed to explain the observed behavior. This is in line with the results of the log-likelihood ratio test. External sources of photons or high-energy particle populations could drive the observed rapid VHE activity as well as the enhanced X-ray activity, as proposed in section \ref{sec:VERFLARE}. Although this is difficult to test given the lack of well-sampled observations, the first detection of a long-duration VHE flaring epoch has provided a test-bed for emission mechanism models, pushing beyond the previously observed rapid variability coming from this source. Future work could include expanding beyond the simple one-zone SSC model to include multi-zone SSC model tests as well as more complex SSC+EIC model interpretations. 

\begin{table}[h]
\caption{Best fit parameters with corresponding $1\sigma$ confidence level range of the one-zone SSC model and the SSC+ EIC model. These modeling results are shown in Figure~\ref{fig:SED}.} 
\label{tab:table1}
\hspace*{-1.25cm}
\begin{tabular}{lcccccr} 
\hline
Parameter & SSC Best Fit &$1\sigma$ range & SSC+EIC Best Fit &  $1\sigma$ range \cr
\hline
 $\delta$ &$70.5$  &[$68.2$, $88.3$] &$89.9$ &[$77.2$, $96.7$]\cr
 $K$ $[\text{cm}^{-3}]$  &$8.7 \times 10^{3}$  &[$8.1 \times 10^{3}$, $1.1 \times 10^{4}$] & $1.6 \times 10^{4}$ & [$9.0 \times 10^{3}$, $2.3 \times 10^{4}$]\cr
 $n_{1}$  &2.16  &[2.14, 2.18] &2.33 & [2.23, 2.34]\cr
 $n_{2}$  &4.37  &[4.32, 4.40] &4.39 & [4.29, 4.39] \cr
 $\gamma_{\text{min}}$  &4.2   &[1.1, 11.0] & 1.7 & [1.7, 18.2]\cr
 $\gamma_{\text{max}}$  &$1.3 \times 10^{6}$  &[$9.7 \times 10^{5}$, $4.6 \times 10^{7}$] & $3.7 \times 10^{7}$ & [$2.7 \times 10^{6}$, $6.2 \times 10^{7}$]\cr
 $\gamma_{\text{break}}$  &$1.8 \times 10^{4}$  &[$1.6 \times 10^{4}$, $1.9 \times 10^{4}$] & $8.0 \times 10^{3}$ & [$6.6 \times 10^{3}$, $8.3 \times 10^{4}$]\cr
 $B$ $[G]$ & $8.4 \times 10^{-3}$ &[$6.5 \times 10^{-3}$, $9.0 \times 10^{-3}$] & $1.7 \times 10^{-2}$ & [$ 1.6\times 10^{-2}$, $2.2 \times 10^{-2}$]\cr
 $ R $ $[\text{cm}]$ &$3.7 \times 10^{16}$ &[$3.2 \times 10^{16}$, $3.8 \times 10^{16}$]   &$ 3.1 \times 10^{16}$ & [$2.1 \times 10^{16}$, $4.0 \times 10^{16}$] \cr
 $bb_{\text{temp}}$ $[K]$ & -- &  -- & $5.9 \times 10^{4}$ & [$ 4.6\times 10^{4}$, $8.2 \times 10^{4}$] \cr
 $L_{\text{nuc}}$ $[\text{erg s}^{-1}]$ & -- & --& $1.6 \times 10^{44}$ & [$ 4.2\times 10^{42}$, $3.4 \times 10^{44}$] \cr
 $\tau$ & -- &  --& $2.4 \times 10^{-4}$ & [$ 9.6\times 10^{-5}$, $4.7 \times 10^{-1}$] \cr
 $D_{\text{BH}}$ $[\text{cm}]$ & -- & -- & $4.5 \times 10^{17}$ & [$ 3.5\times 10^{17}$, $3.5 \times 10^{18}$] \cr
 \hline
\end{tabular}
\end{table}

\section{Discussion \& Conclusion}

The results of a multiwavelength observation campaign of BL\,Lacertae triggered following the report of elevated GeV activity detected by \textit{Fermi}-LAT on October 15, 2022 UTC (\citealt{2022ATel15688....1L}) are presented. For the first time, BL\,Lacertae has been observed in a long-duration VHE flaring state, lasting until at least $\sim$\,40 days after the initial elevated GeV activity was detected. This first long-duration VHE flare detection provides a test-bed of the emission mechanisms responsible for the observed emission originating from IBLs. 

In addition to the observed elevated GeV activity on October 15, 2022 UTC and the long-duration VHE activity detected with VERITAS, multiple epochs of elevated multiwavelength activity during the epoch of interest (September~1, 2022 to December~1, 2022 UTC) are identified via visual inspection. The results of a Local Cross Correlation Function indicated near perfect correlation between the \textit{Fermi}-LAT and ATLAS lightcurves during the GeV flare epoch, suggesting a co-spatial production site. This behavior can be explained by a single-zone SSC process, but could not be tested due to the lack of multiwavelength coverage during this epoch. Following this GeV flare, on October 29, 2022 UTC, VERITAS measured the highest VHE flux from BL\,Lacertae, with minor optical brightening identified via visual inspection. Given the lack of multiwavelength activity, this suggests a more complex scenario as compared to the single-zone SSC model previously proposed. This could allude to a more complex jet structure with multiple SSC zones or relativistic particle populations. 

While VERITAS paused observations of the source following this enhanced VHE activity due to moonlight conditions, elevated X-ray activity is noted on November 12, 2022 UTC with no major multiwavelength counterparts. This could suggest an injection of a new high-energy particle population leading to a hardening of the electron distribution. If SSC mechanisms are assumed in this case, these X-ray photons act as seed photons for IC processes, which would then be reflected in the VHE regime. Unfortunately, as noted previously, VERITAS paused observations of the source during this time period due to moonlight conditions, so no VHE counterpart could be confirmed. However, once VERITAS resumed observations starting November 17, 2022 UTC, strong long-sustaining VHE activity was observed from the source, lasting at least 11 days. This is the first long-duration VHE flare observed from this source, providing a unique opportunity to perform spectral modeling tests to probe potential emission mechanisms proposed to explain the unique multiwavelength behavior.

Given this is the first detection of a long-duration VHE flare from BL\,Lacertae, we consider both a simple one-zone SSC model, as well as a SSC+EIC model with a thermal external inverse-Compton (EIC) component as a first attempt to test possible emission mechanisms responsible for the observed emission. While both models describe the data well, the SSC+EIC model is preferred over the pure SSC model with a significance of $6.9\sigma$. This significant statistical preference suggests that SSC+EIC mechanisms, rather than a one-zone SSC model, are preferred to explain the observed VHE emission from BL Lacertae during this elevated VHE activity epoch. Given the overall complexity of the multiwavelength behavior during the entire epoch of interest in conjunction with these modeling results, this suggests that a more complex model beyond the simple one-zone SSC model is needed to explain the observations. Future work could include expanding beyond the simple one-zone SSC model to include multi-zone SSC model tests as well as more complex SSC+EIC model interpretations. Furthermore, continuing studies on BL\,Lacertae and other IBLs in both flaring and non-flaring states are needed to determine which emission mechanism is dominantly responsible for their emission, as well as to test if this emission mechanism changes during periods of elevated activity.

\section{Acknowledgments}

This research is supported by grants from the U.S. Department of Energy Office of Science, the U.S. National Science Foundation and the Smithsonian Institution, by NSERC in Canada, and by the Helmholtz Association in Germany. This research used resources provided by the Open Science Grid, which is supported by the National Science Foundation and the U.S. Department of Energy's Office of Science, and resources of the National Energy Research Scientific Computing Center (NERSC), a U.S. Department of Energy Office of Science User Facility operated under Contract No. DE-AC02-05CH11231. We acknowledge the excellent work of the technical support staff at the Fred Lawrence Whipple Observatory and at the collaborating institutions in the construction and operation of the instrument.

This work has made use of data from the Asteroid Terrestrial-impact Last Alert System (ATLAS) project. The Asteroid Terrestrial-impact Last Alert System (ATLAS) project is primarily funded to search for near earth asteroids through NASA grants NN12AR55G, 80NSSC18K0284, and 80NSSC18K1575; byproducts of the NEO search include images and catalogs from the survey area. This work was partially funded by Kepler/K2 grant J1944/80NSSC19K0112 and HST GO-15889, and STFC grants ST/T000198/1 and ST/S006109/1. The ATLAS science products have been made possible through the contributions of the University of Hawaii Institute for Astronomy, the Queen’s University Belfast, the Space Telescope Science Institute, the South African Astronomical Observatory, and The Millennium Institute of Astrophysics (MAS), Chile.

This study makes use of VLBA data from the VLBA-BU Blazar Monitoring Program (BEAM-ME), funded by NASA through the Fermi Guest Investigator Program 80NSSC23K1508.  
The VLBA is an instrument of the National Radio Astronomy Observatory. The National Radio Astronomy Observatory is a facility of the National Science Foundation operated by Associated Universities, Inc.

This work made use of data supplied by the UK Swift Science Data Centre at the
University of Leicester.

This research has made use of data from the OVRO 40-m monitoring program \citep{2011ApJS..194...29R}, supported by private funding from the California Institute of Technology and the Max Planck Institute for Radio Astronomy, and by NASA grants NNX08AW31G, NNX11A043G, and NNX14AQ89G and NSF grants AST-0808050 and AST-1109911.

C.E.H. received support for this work from the National Science Foundation under Grant No. 2125733.

W.M. gratefully acknowledges support by the ANID BASAL project FB210003.

\bibliography{citations}
\bibliographystyle{aasjournal}

\section*{Appendix}

\begin{figure*}[h]
    \centering
    \includegraphics[width = 7in]{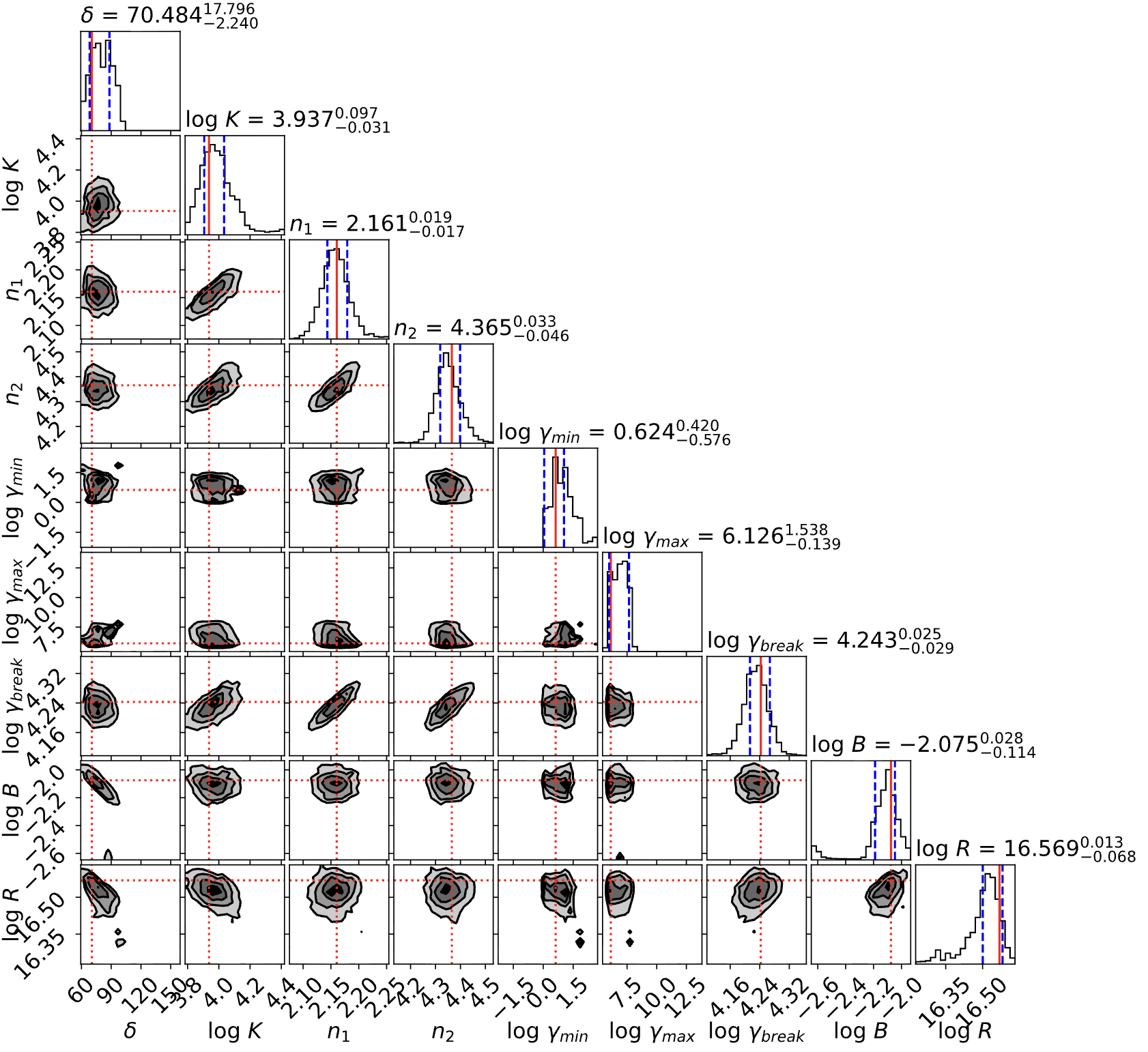}
    \caption{The corner plot of the posterior probability distribution of the free parameters
from the one-zone SSC SED fit.}
    \label{fig:cornerSSC}
\end{figure*}

\begin{figure*}[h]
    \centering
    \includegraphics[width = 7in]{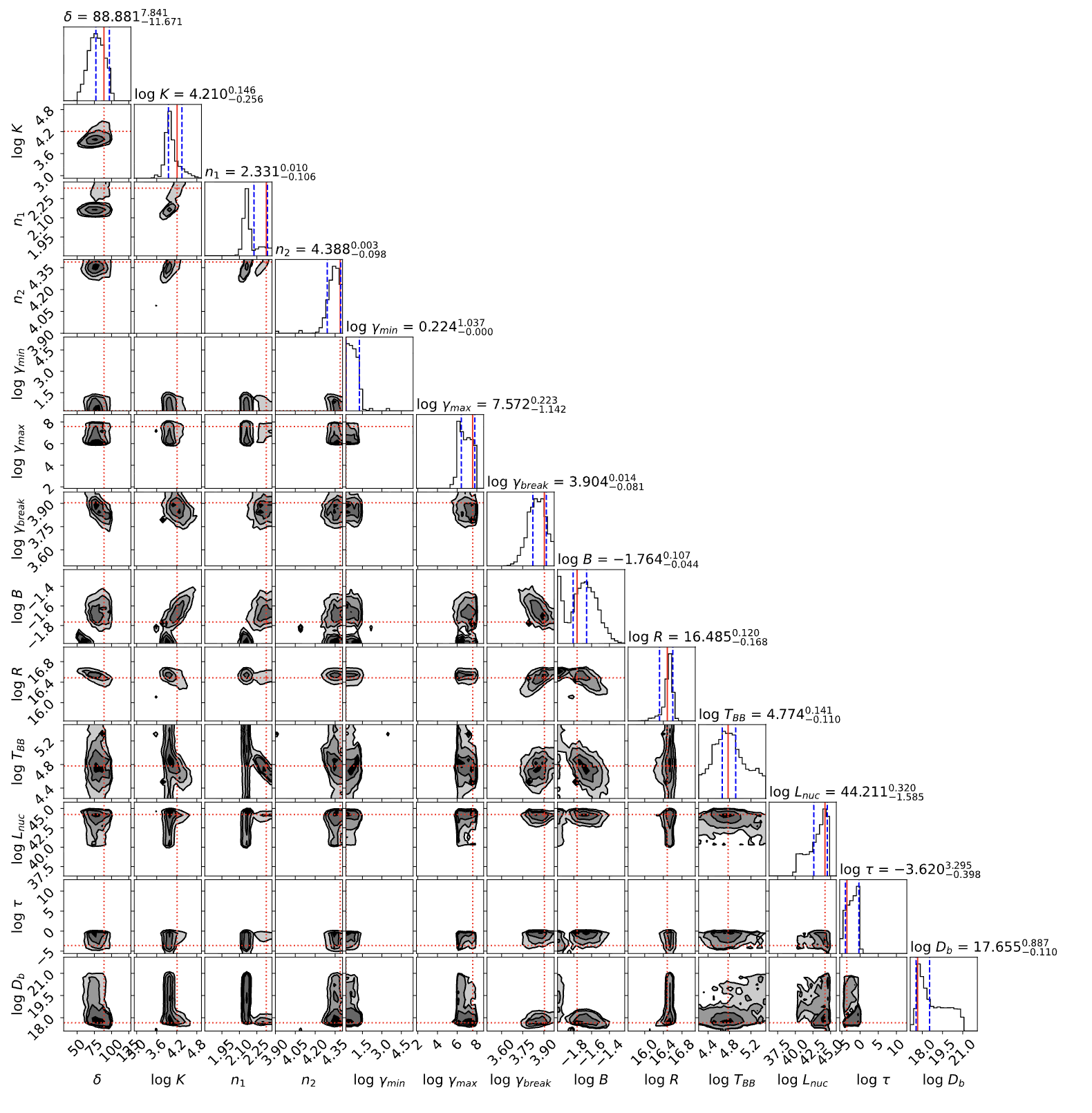}
    \caption{The corner plot of the posterior probability distribution of the free parameters
from the SSC+EIC SED fit.}
    \label{fig:cornerSSC+EIC}
\end{figure*}

\end{document}